Considerations on the motion of celestial bodies
*E304 -- Considerationes de motu corporum coelestium*

Originally published in *Novi Commentarii academiae scientiarum Petropolitanae* 10, 1766, pp. 544-558.
Also published in *Opera Omnia*: Series 2, Volume 25, pp. 246 – 257.
According to the records, a treatise with this title was read to the Berlin Academy on April 22, 1762, and it was presented to the Petersburg Academy on May 17, 1762.

Translated and Annotated[†]
(Draft)
by
Sylvio R. Bistafa[*]

May 2019

Foreword

Euler wrote a number of papers in Astronomy, most of them in Latin. This is a commented translation of *E304 -- Considerationes de motu corporum coelestium* (Considerations on the motion of celestial bodies). In this publication, Euler essentially focuses on the solution of two particular motions of a three-body problem consisting of Sun, Earth and Moon. The first motion, represents a hypothetical situation of these three celestial bodies in perpetual alignment in syzygy — the three-body problem on a straight line —, for which the parameter that controls the distances among the planets was found to be given by a quintic function. The conclusion was that if the Moon were four times more distant from the Earth (either in conjunction or in opposition), a motion of this kind would have been possible to exist, such that the Moon would appear always connected to the Sun. The second motion considered by Euler was Moon libration, when these planets are aligned in regular syzygy. Here, perhaps for the first time, Euler introduces an archaic form of a Fourier sine series expansion to describe the Moon's wagging motion. However, as Euler himself recognizes, the calculations turned out very tedious, and led him to greatly simplify his model in order to obtain some numerical values for the phenomenon.

Considerations on the motion of celestial bodies
Author
L. Euler

I.

Although there is no doubt that the laws of motion of celestial bodies observed by Kepler and confirmed by Newton have brought very great gains to [the discipline of] Astronomy, nevertheless it is certain that no body in the heavens is met with that in its own motion follows these laws perfectly, since, instead, in all [of the motions of these bodies] deviations from these laws, that are by no means slight, are detected. Of course, it is true that the cause of all heavenly motions resides in the mutual attraction of these bodies, by which each and every [body] is attracted toward each of the others singly by forces consisting of a ratio composed directly by the un-squared [amount] of the masses [of the bodies], and inversely by the squared [amounts] of the distances. However, it is always convenient to consider that one force stands out among the remaining, and thus, the motion would approximately follows Kepler's rules; and thus the relatively very small effect arising from the others can be determined by methods of approximation. Without this simplification, we would be at the utmost ignorance about the celestial motion, since to date no method has been discovered by the application of which the motion, of three or more bodies mutually attracting each other, might require to be ascribed; unless, perchance, one force surpasses the others.

---

[†] The translator used the best of his abilities and knowledge to make this translation technically and grammatically as sound as possible. Nonetheless, interested readers are encouraged to make suggestions for corrections as they see fitting.
[*] Corresponding address: sbistafa@usp.br

2. Yet indeed this case—in which [case] alone Geometers do not squander their work at all points in vain — cannot be taken as conclusive since the method of approximation itself, which Geometers are accustomed to use, is bound up with a great many difficulties besides, and an unlimited multitude of small perturbations is neglected, by which [fact] it becomes so that this approximation by itself [only] minimally carries through the business [of determining the motions], but on the contrary, for it to be completed, still more supports are desired. Wherefore, although from this Theory the motion of the Moon is determined accurately enough, that [fact of sufficient accuracy] ought to be ascribed more to special circumstances that obtain for the Moon than to any perfection to which [perfection] a general Theory would be required to measure up. For if the Moon were two or three times as far from the Earth, or if its orbit were more eccentric, then all the labors endured to this point would be lacking in all fruit [because they would be inapplicable], and by the way, not even its motion could be recalled to any fixed rule.

3. Therefore, much have stood before the Theory of Astronomy to be considered, for instance, if under the fictitious hypothesis that in case the Moon were much away from the earth, it would be certainly an excess to think that its motion could be evaluated with the maximum aid of this science. If, for instance, the Moon would have been a hundred times more distant from the earth, there is no doubt that the laws of motion of the main planet would no longer be followed as if it were a satellite of the earth, as one would expect. But if, on the other hand, the distance were ten times greater, its motion could then be compared, so that no doubt would remain, even with primary or secondary planets being added. To such an extent that it certainly would disagree from all the motions observed in the sky, such that it can hardly give an idea even on how the average motion can be resolved. Perhaps, innumerable observations could reveal a certain law, from which, in a subsequent application, it can somehow give a clear prediction; however, by no means so evident, to such an extent that the Theory that should explain this type of motion may not be adapted. The very wise creator is seen to have being mindful of our weakness, because none of the bodies placed in the sky are such that their motion could be described neither by the law of the main planets nor of the satellites.

4. This sort of research, which is seen not to surpass the strength of the human mind, is certainly not suited to be undertaken hastily, but on the contrary, it will require that our efforts be undertaken step by step. Then, the general problem of three bodies mutually attracted to each other will be conveniently restricted to the case where one of the masses almost vanishes in front of the two remaining, where it is agreed that certainly it will be convenient that the two larger bodies are set in motion according to Kepler's laws, and that every perturbation on the third [body] is disconsidered, and in case its position and motion will be compared since the beginning, such that if it is attracted to both larger [bodies] with almost equal force, we shall have in this manner a case, whose investigation demands a distinctly new approach. A great deal is lacking to venture an approach towards this problem before fatiguing in vain to unfold it, as I am forced to admit; however, in fact, this is a complete singular case as I have already observed, and with a remarkable simplification, in which the motion of the Moon would appear constantly connected or in opposition to the Sun, which is the case to be considered, with great utility in this very difficult matter should not be abandoned, and by no means to be seen with indifference.

5. Hence, the motion of the Sun and the Moon seen from Earth is assumed to take place in the ecliptic plane[1], with the earth resting in $T$, and after a certain time has elapsed, I place the Sun in $S$, the Moon in $L$, and after laying a fixed straight line $TA$, directed to the First Star of Aries[2], I ascribe the following angles: $ATS = \theta$, $ATL = \phi$, and $STL = \phi - \theta = \eta$, and the distances $TS = u$, $TL = v$ and $LS = \sqrt{uu - 2uv\cos\eta + vv} = z$. Be further the mean longitude of the Sun[3] $= \zeta$, and its mean distance from the Earth $= a$, and from these we have for the motion of the Sun in its canonical form[4]:

$$\frac{2du d\theta + u dd\theta}{d\zeta^2} = 0 \text{ and } \frac{ddu - u d\theta^2}{d\zeta^2} + \frac{a^3}{u^2} = 0,$$

and for the motion of the moon[5]:

$$\frac{2dv d\phi + v dd\phi}{d\zeta^2} - \frac{a^3}{u^2}\left(1 - \frac{u^3}{z^3}\right)\sin\eta = 0, \text{ and}$$

$$\frac{ddv - v d\phi^2}{d\zeta^2} + \frac{n^2 c^3}{v^2} + \frac{a^3 v}{z^3} + \frac{a^3}{u^2}\left(1 - \frac{u^3}{z^3}\right)\cos\eta = 0$$

where $c$ is the mean distance which the Moon is solicited by the force of the Earth, and for the mean motion of revolution, there is a $n:1$ relation between the mean motion of the Moon and the mean motion of the Sun. Besides, regarding the differentials of the second degree, it should be noticed that the element $d\zeta$ is assumed to be constant.

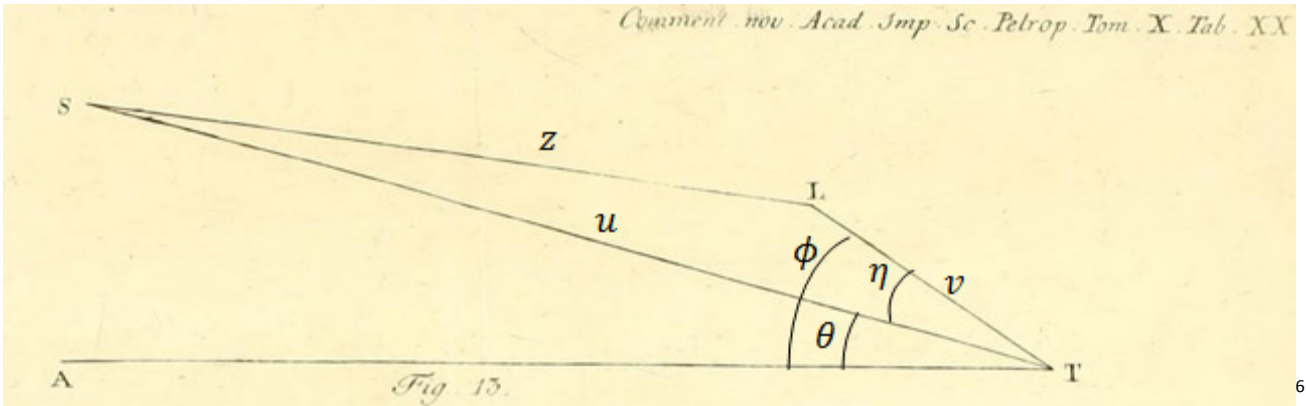

6. All the difficulties in the resolution of both equations, consist on finding at any instant of time, the mean longitude of the Sun $\zeta$, as well as the distance $v$, and the angle $\phi$. Since so far, in general, the Geometers cannot proceed in their work, unless in case in which the distance $v$ is much smaller than $u$ and the number $n$ is rather big, suitable approximations have been found, yet much is still justly needed in this matter. However this pair of equations in a general aspect, without any consideration to the Moon dwelling, which certainly has to be explained, can solve the problem completely. Such motion can take place in the sky, and it is in our ability to know it completely, even if its reasoning does not agree at all with the regular motion.

---

[1] The ecliptic plane contains most of the objects which are orbiting the sun, and is tilted with respect to the Earth's spin axis at 23.5°.
[2] The First Star of Aries (or First Point of Aries), also known as the Cusp of Aries, is the location of the vernal equinox.
[3] The Sun's ecliptical longitude is defined as the angle subtended at the earth between the vernal equinox and the Sun. The mean longitude is the ecliptical longitude that the planet would have if the orbit were a perfect circle.
[4] The development of these equations can be found in E112 *Recherches sur le mouvement des corps célestes en general*.
[5] The development of these equations can be found in L. Euler, *Considerationes de theoria motus lunae perficienda et imprimis de eius variatione, Novi Commentarii Academie Scientiarum Imperialis Petropolitanae, Tom. XIII, pro Anno 1768*.
[6] The angles were added to this figure by the Translator to facilitate the comprehension.

7. First of all, I should observe that these two equations admit a closed form solution for the case in which $\eta = 0$, or $\phi = \theta$, when the Moon is seen to be in continuous communication with the Sun. Since $sin\,\eta = 0$, and $cos\,\eta = 1$, then $z = u - v$, and our equations will assume the following forms:

$$\frac{2dvd\theta + vdd\theta}{d\zeta^2} = 0, \text{and}$$

$$\frac{ddv - vd\theta^2}{d\zeta^2} + \frac{n^2c^3}{v^2} + \frac{a^3v}{(u-v)^3} + \frac{a^3}{u^2} \cdot \frac{-3u^2v + 3uv^2 - v^3}{(u-v)^3} = 0$$

$$\text{or } \frac{ddv - vd\theta^2}{d\zeta^2} + \frac{n^2c^3}{v^2} + \frac{a^3v(2u^2 - 3uv + v^2)}{u^2(u-v)^3} = 0,$$

which can be immediately compared with the given formulas for the motion of the Sun considering that $v = \alpha u$, which it is certainly satisfied by the prior established equations. Hence the other equation for the Moon will be transformed into

$$\frac{\alpha(ddu - ud\theta^2)}{d\zeta^2} + \frac{n^2c^3}{\alpha^2 u^2} + \frac{\alpha a^3(2 - 3\alpha + \alpha^2)}{(1-\alpha)^3 u^2} = 0.$$

And since the other equation for the Sun is

$$\frac{ddu - ud\theta^2}{d\zeta^2} + \frac{a^3}{u^2} = 0,$$

it is necessary that

$$\alpha a^3 = \frac{n^2c^3}{\alpha^2} - \frac{\alpha a^3(2 - 3\alpha + \alpha^2)}{(1-\alpha)^3}$$

or

$$\frac{n^2c^3}{\alpha^2 a^3} - \frac{3\alpha - 3\alpha^2 + \alpha^3}{(1-\alpha)^2},$$

where, since $\frac{n^2c^3}{a^3}$ is a constant quantity, let us put for conciseness $\frac{n^2c^3}{a^3} = m$, and then

$$m(1-\alpha)^2 = \alpha^2(3\alpha - 3\alpha^2 + \alpha^3)$$

or

$$m(1-\alpha)^2 = \alpha^2 - \alpha^2(1-\alpha)^3.$$

Once putted $2 - \alpha = x$, then $mx^2 = (1 - x^3)(1 - x)^2$, or

$$1 - 2x + x^2 - mx^2 - x^3 + 2x^4 - x^5 = 0.\text{[7]}$$

---

[7] A fifth degree polynomial was also obtained by Euler in E327 -- *De motu rectilineo trium corporum se mutuo attrahentium*. In this publication, Euler considers three bodies laying on a straight line, which are attracted to each other by central forces inversely proportional to the square of their separation distance (inverse-square law).

8. Then, it will be necessary the determination of the number $\alpha$ or $x$ from an equation of the fifth degree, which for its resolution, it is necessary to first observe that $m$ should be a rather small fraction, and then

$$m(1-\alpha)^2 = 3\alpha^3 - 3\alpha^4 + \alpha^5,$$

likewise, it is evident that $\alpha$ will also have a small value, which can be approximated by $\alpha = \sqrt[3]{\frac{m}{3}} = \frac{c}{a}\sqrt[3]{\frac{n^2}{3}}$, or more precisely $\alpha = \sqrt[3]{\frac{m}{3}} - \frac{1}{3}\sqrt[3]{\frac{m^2}{9}} - \frac{1}{27}m + \frac{1}{81}m\sqrt[3]{\frac{m}{3}}$. However, a first approximation gives $v = \frac{cu}{a}\sqrt[3]{\frac{n^2}{3}}$, whence, since $u = a$, and $n^2 = 175$,[8] gives, $v = 4c$, approximately; or if the Moon were four times more distant from us, a motion of this kind would have been possible to exist, such that it [the Moon] would appear always connected to the Sun. It would then be possible to regard a Satellite of the Earth as if it were the Moon, and its motion would be most regular, however, deviating from the rules of Kepler the more close to the Sun than to the Earth, though it may revolved with the same time, because the force of the Earth in relation to the force of the Sun is reduced in the same proportion, although it may linger with a longer periodic time. Because the distance to the earth would be almost four times greater than the distance that Moon actually stands apart, as much as a limit would permit, so that bodies far more removed from the principal planets, in fact closer to the satellites of the Earth should permit. Similar limits in relation to other planets will be possible to be established.

9. As the evolved case of a permanent continuous communication with the Sun, indeed in continuous opposition, gives a similar case. In this case, let us put $\eta = 180°$, and then $\sin\eta = 0$, $\cos\eta = -1$, and $\phi = 180° + \theta$, and then $d\phi = d\theta$, and also $z = u + v$. Thus, the equations for the motion of the Moon will assume the following forms:

$$\frac{2dvd\theta + vdd\theta}{d\zeta^2} = 0, \text{and}$$

$$\frac{ddv - vd\theta^2}{d\zeta^2} + \frac{n^2c^3}{v^2} + \frac{a^3v}{(u+v)^3} - \frac{a^3}{u^2}\left(1 - \frac{u^3}{(u+v)^3}\right) = 0,$$

and the latter one reducing to the following:

$$\frac{ddv - vd\theta^2}{d\zeta^2} + \frac{n^2c^3}{v^2} - \frac{a^3}{u^2} - \frac{a^3}{(u+v)^2} = 0.$$

First, considering also the motion of the Sun, gives at once $v = \alpha u$, which transform this equation into

$$\frac{\alpha(ddu - ud\theta^2)}{d\zeta^2} + \frac{n^2c^3}{\alpha^2 u^2} - \frac{a^3}{u^2} - \frac{a^3}{(1+\alpha)^2 u^2} = 0.$$

Yet, for the motion $\frac{ddu - ud\theta^2}{d\zeta^2} = -\frac{a^3}{u^2}$, which transforms the equation into

$$-\alpha a^3 + \frac{n^2c^3}{\alpha^2} - a^3 + \frac{a^3}{(1+\alpha)^2} = 0, \text{or}$$

---

[8] The Moon completes an orbit around the Earth once every 27.32 days. The Earth takes a year (365 days) to revolve around the Sun. Therefore, in a year period, the Moon completes $n = 365/27{,}32 = 13{,}36$ revolutions around the Earth, and then $n^2 = 178{,}5$.

$$\frac{n^2c^3}{a^3} - \alpha^2(1+\alpha) + \frac{\alpha^2}{(1+\alpha)^2} = 0$$

and once put, for conciseness, $\frac{n^2c^3}{a^3} = m$, gives

$$m(1+\alpha)^2 = \alpha^2(1-\alpha)^3 - \alpha^2,$$

which is obtained from the equation of the same form above by taking $m$ and $\alpha$ negative.

Henceforth

$$\alpha = \sqrt[3]{\frac{m}{3}} + \frac{1}{3}\sqrt[3]{\frac{m^2}{9}} - \frac{1}{27}m - \frac{1}{81}m\sqrt[3]{\frac{m}{3}},$$

however, as a first approximation $\alpha = \sqrt[3]{\frac{m}{3}}$ and $v = \frac{cu}{a}\sqrt[3]{\frac{n^2}{3}}$ as before.

10. These cases are most worthy to be commented, since they could be worked out absolutely without any approximation, even if both forces of the Sun and of the Earth concur in producing motion, since there is no other case which this can happen. However, the body would, in fact, move with such a simple motion, provided it would be at the assigned distance, and while it would appear from the Earth in conjunction or in opposition with the Sun, a motion of this type would be impressed, when it had began to advance in the same pace with the Earth in the ecliptic plane. However, on the other hand, if the impressed motion differs from this law, it would not, in fact, remain in continuous conjunction or in opposition with the Sun, but it would perform tiny excursions and, because of this, as almost oscillating. In the case where the motion had minimally differed from the formulation that was found, in the usual way, also by approximation, it will be possible to define such motion; in this case with the threshold of irregular motions, which still cannot be approached by any calculation, certainly it is seen with no lack of usefulness, if I will seek more carefully the nature of such motion.

11. However, although this investigation is by no means involved by trivial difficulties, however, our equations can be made conveniently easier to handle, when the distance $v$ is much smaller than $u$, when it is possible to produce a convenient approximation. Of course, because $z = \sqrt{uu - 2uv\cos\eta + vv}$, we have approximately $\frac{1}{z^3} = \frac{1}{u^3} + \frac{3v\cos\eta}{u^4} - \frac{3v^2}{2u^5} + \frac{15v^2\cos\eta^2}{2u^5}$, and thus $1 - \frac{u^3}{z^3} = -\frac{3v}{u}\cos\eta - \frac{3v^2}{2u^2}(1 - 5\cos\eta^2)$, from which our equations that were found for the motion of the Moon will transform into the following forms:

$$I. \frac{2dvd\phi + vdd\phi}{d\zeta^2} - \frac{3a^3v}{u^3}\sin\eta\cos\eta - \frac{3a^3v^2}{2u^4}\sin\eta(1 - 5\cos\eta^2) = 0$$

$$II. \frac{ddv - vd\phi^2}{d\zeta^2} + \frac{n^2c^3}{v^2} + \frac{a^3v}{z^3}\left(1 - 3\cos\eta^2 + \frac{3a^3v^2}{2u^4}(3\cos\eta - 5\cos\eta^3)\right) = 0.$$

Then, also, the calculation can be made easier, if we consider the mean motion of the Sun, then $u = a$, and $\theta = \zeta$, and thus $\eta = \phi - \zeta$, or $\phi = \eta + \zeta$, whence arising the following equations

$$I. \frac{2dvd\eta + vdd\eta}{d\zeta^2} + \frac{2dv}{d\zeta} + 3v\sin\eta\cos\eta - \frac{3v^2}{2a}\sin\eta(1 - 5\cos\eta^2) = 0$$

$$II. \frac{ddv}{d\zeta^2} - v\left(1 + \frac{d\eta}{d\zeta}\right)^2 + v(1 - 3\cos\eta^2) + \frac{n^2c^3}{v^2} + \frac{3v^2}{2a}\cos\eta(3 - 5\cos\eta^2) = 0,$$

where the last terms in these expressions can be omitted, since the fraction $\frac{v}{a}$ is very small, even establishing that the distance of the Moon is four times larger.

12. Now, be reminded the case where the Moon will be seen hesitating in a almost oscillating motion around the Sun, and let us assume that the angle $\eta$ is as small as possible, such that $\sin\eta = \eta$, and $\cos\eta = 1 - \frac{1}{2}\eta^2$, and then we have:

$$I. \frac{2dvd\eta + vdd\eta}{d\zeta^2} + \frac{2dv}{d\zeta} + 3v\eta = 0$$

$$II. \frac{ddv}{d\zeta^2} - v\left(1 + \frac{d\eta}{d\zeta}\right)^2 + \frac{n^2c^3}{v^2} - 2v + 3v\eta^2 = 0.$$

Then, because the distance $v$ is little changed, let us put $v = b(1+x)$, such that $x$ is a small quantity, and further, be for brevity $\frac{n^2c^3}{b^3} = m$, and hence

$$I. \frac{2dxd\eta + xdd\eta}{d\zeta^2} + \frac{dd\eta}{d\zeta^2} + \frac{2dx}{d\zeta} + 3\eta + 3x\eta = 0$$

$$II. \frac{ddx}{d\zeta^2} - 3 - 3x - \frac{2d\eta}{d\zeta} - \frac{2xd\eta}{d\zeta} - \frac{d\eta^2}{d\zeta^2} - \frac{xd\eta^2}{d\zeta^2} + 3\eta^2 + 3x\eta^2 + m - 2mx + 3mx^2 = 0.$$

whence, it is necessary to define the values of the quantities $x$ and $\eta$ for every angle $\zeta$.

13. Since the angle $\eta$ is minimum, and which alternates between negative and positive values, as the Moon is seen passing to and fro the Sun: it is easily allowed to conceive the existence of a relation between a certain angle $\omega$ and $\zeta$, and thus to define the following

$$\eta = A\sin\omega + B\sin2\omega + C\sin3\omega \text{ etc.}\ ^9$$

and also $d\omega = \alpha d\zeta$. Then,

$$\frac{d\eta}{d\zeta} = \alpha A\cos\omega + 2\alpha B\cos2\omega + 3\alpha C\cos3\omega, \text{ and}$$

$$\frac{dd\eta}{d\zeta^2} = -\alpha^2 A\sin\omega - 4\alpha^2 B\sin2\omega - 9\alpha^2 C\sin3\omega.$$

On account that the first equation can be transformed into

$$\frac{2dx}{1+x} + \frac{dd\eta + 3\eta d\zeta^2}{d\eta + d\zeta} = 0$$

which, upon integration gives

$$2\ln(1+x) + \ln\left(1 + \frac{d\eta}{d\zeta}\right) + 3\int \frac{\eta d\zeta}{1 + \frac{d\eta}{d\zeta}} = Const.$$

or, because $x$ and $\frac{d\eta}{d\zeta}$ are small quantities, then:

$$2x - x^2 + \frac{2}{3}x^3 + \frac{d\eta}{d\zeta} - \frac{d\eta^2}{2d\zeta^2} + \frac{d\eta^3}{3d\zeta^3} + 3\int \eta d\zeta - \frac{3}{2}\eta^2 + 3\int \frac{\eta d\eta^2}{d\zeta} - 3\int \frac{\eta d\eta^3}{d\zeta^2} = Const.\ ^{10}$$

---

[9] This appears to be an archaic form of a Fourier sine series expansion.

Now, since $d\zeta = \frac{d\omega}{\alpha}$, then

$$\int \eta d\zeta = -\frac{A}{\alpha}\cos\omega - \frac{B}{2\alpha}\cos2\omega - \frac{C}{3\alpha}\cos3\omega$$

$$\eta^2 = \frac{1}{2}A^2 + AB\cos\omega - \frac{1}{2}A^2\cos2\omega - AB\cos3\omega$$
$$+\frac{1}{2}B^2 \qquad\qquad + AC$$

$$\frac{d\eta^2}{d\zeta^2} = \frac{1}{2}\alpha^2 A^2 + 2\alpha^2 AB\cos\omega + \frac{1}{2}\alpha^2 A^2\cos2\omega + 2\alpha^2 AB\cos3\omega$$
$$+2\alpha^2 B^2 \qquad\qquad + 3\alpha^2 AC$$

$$\frac{d\eta^3}{d\zeta^3} = \alpha^3 A^2 B + \frac{3}{4}\alpha^3 A^3 \cos\omega + \alpha^3 A^2 B\cos2\omega$$
$$+4\alpha^3 AB^2,$$

where we justly disregarded higher powers of the letters $A, B, C$.

14. Since

$$\frac{\eta d\eta^2}{d\zeta^2} = \frac{1}{4}\alpha^2 A^3 \sin\omega + \frac{3}{2}\alpha^2 A^2 B\sin2\omega + \frac{1}{4}\alpha^2 A^3 \sin3\omega$$
$$+3\alpha^2 AB^2 \quad + 2\alpha^2 B^3 \qquad +\frac{3}{2}\alpha^2 A^2 C$$
$$-\frac{3}{2}\alpha^2 A^2 C \qquad\qquad\qquad + \alpha^2 AB^2,$$

and because $d\zeta = \frac{d\omega}{\alpha}$, then

$$\int \frac{\eta d\eta^2}{d\zeta} = -\frac{1}{4}\alpha A^3 \cos\omega - \frac{3}{4}\alpha A^2 B\cos2\omega - \frac{1}{12}\alpha A^3 \cos3\omega$$
$$-3\alpha AB^2 \quad - \alpha B^3 \qquad -\frac{1}{2}\alpha A^2 C$$
$$+\frac{3}{2}\alpha A^2 C \qquad\qquad\qquad -\frac{1}{2}\alpha AB^2,$$

where, since the series $A, B, C$ had already decreased very much, further terms can be omitted. Then, since

$$\frac{\eta d\eta^3}{d\zeta^3} = \frac{7}{8}\alpha^3 A^3 B\sin\omega + \frac{3}{8}\alpha^3 A^4 \sin2\omega + \frac{7}{8}\alpha^3 A^3 B\sin3\omega$$

Which upon integration gives

---

[10] These integrals are the result of a series expansion of $\frac{1}{1+\frac{d\eta}{d\zeta}} = 1 - \frac{d\eta}{d\zeta} + \frac{d\eta^2}{d\zeta^2} - \frac{d\eta^3}{d\zeta^3}$, and then, $3\int\frac{\eta d\zeta}{1+\frac{d\eta}{d\zeta}} = 3\int \eta d\zeta - 3\int \eta d\eta + 3\int\frac{\eta d\eta^2}{d\zeta} - 3\int\frac{\eta d\eta^3}{d\zeta^2}$.

$$\int \frac{\eta \, d\eta^3}{d\zeta^2} = -\frac{7}{8}\alpha^3 A^3 B \cos\omega - \frac{3}{16}\alpha^3 A^4 \cos 2\omega - \frac{7}{24}\alpha^3 A^3 B \cos 3\omega,$$

and finally, from the expressions above, and omitting the terms that are constants, the following equation is obtained

$$2x - x^2 + \frac{2}{3}x^3 + \alpha A \cos\omega + 2\alpha B \cos 2\omega + 3\alpha C \cos 3\omega = 0 \ ^{11}$$

$$
\begin{array}{ccc}
-\alpha^2 AB & -\frac{1}{4}\alpha^2 A^2 & -\alpha^2 AB \\
+\frac{1}{4}\alpha^3 A^3 & -\frac{3}{2}\alpha^2 AC & -\frac{C}{\alpha} \\
-\frac{3A}{\alpha} & +\frac{1}{3}\alpha^3 A^2 B & +\frac{3}{2}AB \\
-\frac{3}{2}AB & -\frac{3B}{2\alpha} & -\frac{1}{4}\alpha A^3 \\
-\frac{3}{4}\alpha A^3 & +\frac{3}{4}A^2 & +\frac{1}{4}\alpha^3 A^3 \\
 & -\frac{3}{2}AC & \\
 & -\frac{9}{4}\alpha A^2 B &
\end{array}
$$

15. To find the value of $x$, let us put for conciseness

$$\left(\alpha - \frac{3}{\alpha}\right)A - \left(\alpha^2 + \frac{3}{2}\right)AB + \frac{1}{4}\alpha(\alpha^2 - 3)A^3 = \mathfrak{A}$$

$$\frac{4\alpha^2 - 3}{2\alpha}B - \frac{(\alpha^2 - 3)}{4}A^2 = \mathfrak{B}$$

$$\frac{3\alpha^2 - 1}{\alpha}C - \frac{(2\alpha^2 - 3)}{2}AB + \frac{1}{4}\alpha(\alpha^2 - 1)A^3 = \mathfrak{C}$$

such that

$$2\ln(1+x) + \mathfrak{A}\cos\omega + \mathfrak{B}\cos 2\omega + \mathfrak{C}\cos 3\omega = 0,$$

and then

$$1 + x = e^{-\frac{1}{2}\mathfrak{A}\cos\omega - \frac{1}{2}\mathfrak{B}\cos 2\omega - \frac{1}{2}\mathfrak{C}\cos 3\omega},$$

whence we conclude that

$$
\begin{aligned}
x = &-\frac{1}{2}\mathfrak{A}\cos\omega - \frac{1}{2}\mathfrak{B}\cos 2\omega - \frac{1}{2}\mathfrak{C}\cos 3\omega \\
&+ \frac{1}{8}\mathfrak{A}\mathfrak{B} \quad + \frac{1}{16}\mathfrak{A}^2 \quad + \frac{1}{8}\mathfrak{A}\mathfrak{B} \\
&+ \frac{1}{64}\mathfrak{A}^3 \quad \quad\quad\quad -\frac{1}{192}\mathfrak{A}^3.
\end{aligned}
$$

---

[11] The expanded forms of the integrals $3\int \frac{\eta \, d\eta^2}{d\zeta} - 3\int \frac{\eta \, d\eta^3}{d\zeta^2}$ were not included in this equation.

But for us not to be involved in excessively tedious calculations, we shall procure a less accurate expression, by neglecting the triple angle, such that $\eta = A\sin\omega + B\sin2\omega$, and then we have

$$x = -\frac{(\alpha^2 - 3)}{2\alpha}A\cos\omega - \frac{(4\alpha^2 - 3)}{4\alpha}B\cos2\omega + \frac{3(\alpha^2 - 1)(\alpha^2 - 3)}{16\alpha^2}A^2$$

where, for conciseness, we put

$$x = E\cos\omega + F\cos2\omega$$

such that

$$E = \frac{3 - \alpha^2}{2\alpha}A \text{ and } F = \frac{3 - 4\alpha^2}{4\alpha}B + \frac{3(\alpha^2 - 1)(\alpha^2 - 3)}{16\alpha^2}A^2.$$

16. When these values are substituted in the second equation, we will find out that[12]:

$$\frac{ddx}{d\zeta^2} = \qquad -\alpha^2 E\cos\omega - 4\alpha^2 F\cos2\omega$$

$$-3 - 3x = \quad -3 \quad\quad -3E \quad\quad -3F$$

$$-\frac{2d\eta}{d\zeta} = \quad -\alpha AE \quad -2\alpha A \quad -4\alpha B$$

$$\qquad\qquad\qquad\qquad -2\alpha BE \quad -\alpha AE$$

$$-\frac{2xd\eta}{d\zeta} = \qquad\qquad -\alpha AF$$

$$-\frac{d\eta^2}{d\zeta^2} = -\frac{1}{2}\alpha^2 A^2 \quad +2\alpha^2 AB \quad -\frac{1}{2}\alpha^2 A^2$$

$$3\eta^2 = +\frac{3}{2}A^2 \quad + 3AB\cos\omega \quad -\frac{3}{2}A^2\cos2\omega$$

$$m = +m$$

$$-2mx = \qquad\qquad -2mE \quad -2mF$$

$$+3mx^2 = \frac{3}{2}mE^2 \quad +3mEF \quad +\frac{3}{2}mE^2$$

whence we first conclude that[13]:

$$m\left(1 + \frac{3}{2}E^2\right) = 3 + \alpha AE + \frac{1}{2}\alpha^2 A^2 - \frac{3}{2}A^2 = 3$$

thence

$$m = 3 - \frac{9}{2}E^2$$

---

[12] The reduced form of $-\frac{xd\eta^2}{d\zeta^2}$ from the second equation was not included in the calculations.

[13] The following expression is the result of equating the constant terms to zero.

for the determination of the number $m$ and then the distance $b$.

It is clear that $m = 3$, approximately, and thus $b = c\sqrt[3]{\frac{n^2}{3}}$. We have further that[14]

$$-\alpha^2 E - 9E - 2\alpha A - 2\alpha BE - \alpha AF (+2\alpha^2 + 3)AB + 9EF = 0$$

which, by neglecting terms which cannot be simply reduced to $\frac{E}{A} = \frac{3-\alpha^2}{2\alpha}$, gives

$$(\alpha^2 + 9)(3-\alpha^2) + 4\alpha^2 = 0, \text{ or}$$

$$\alpha^4 + 2\alpha^2 - 27 = 0, \text{ and then, } \alpha^2 = \sqrt{28} - 1.$$

Finally, the third equation gives[15]

$$-(4\alpha^2 + 9)F - 4\alpha B - \frac{3}{2}A^2 AE - \frac{1}{2}\alpha^2 A^2 - \frac{3}{2}A^2 + \frac{9}{2}E^2 = 0,$$

therefore:

$$\left. \begin{array}{c} \dfrac{(4\alpha^2 - 3)(4\alpha^2 + 9)}{4\alpha}B - \dfrac{3(\alpha^2 - 1)(\alpha^2 - 3)(4\alpha^2 + 9)}{16\alpha^2}A^2 \\ -4\alpha B \qquad\qquad + \dfrac{\alpha^2 - 3}{2}A^2 \\ \qquad\qquad\qquad - \dfrac{(\alpha^2 + 3)}{2}A^2 \\ \qquad\qquad\qquad + \dfrac{9(\alpha^2 - 3)^2}{8\alpha^2}A^2 \end{array} \right\} = 0$$

which, after rearranging gives:

$$B(16\alpha^4 + 8\alpha^2 - 27) = \frac{3}{2}A^2\alpha\left(13 - 7\alpha^2 + \frac{3A^2}{2\alpha}\right),$$

or since $27 = \alpha^4 + 2\alpha^2$, then

$$3B(5\alpha^2 + 2) = \frac{3A^2}{2\alpha}(13 - 7\alpha^2 + 2\alpha^4),$$

and, therefore,

$$B = \frac{13 - 7\alpha^2 + 2\alpha^4}{2\alpha(5\alpha^2 + 2)}A^2 = \frac{67 - 11\alpha^2}{2\alpha(5\alpha^2 + 2)}A^2$$

$$F = \frac{291 - 94\alpha^2 - 23\alpha^4}{8\alpha^2(5\alpha^2 + 2)}A^2 = -\frac{165 - 24\alpha^2}{4\alpha^2(5\alpha^2 + 2)}A^2.$$

Then, accordingly, the value of $A$ is at our discretion, which depends on the digressions from the syzygy line[16], which it is proper to assume as being a very small fraction, such that the quadratic terms can be considered of second order, being sufficient only the first terms. Then, for the distance $v = b(1 + x)$, we have that $b = c\sqrt[3]{\frac{n^2}{3}}$, and the angle $\omega$ is defined such that $\omega = \alpha\zeta + \beta$, and since $\alpha^2 = \sqrt{28} - 1$, then, $\alpha^2 = 4.291502$ and $\alpha = 2{,}071594$. Thereupon, putting $\eta = A\sin\omega$ and $v = b\left(1 - \frac{(\alpha^2-3)}{2\alpha}A\cos\omega\right)$ or $v = b(1 - 0{,}311717 A\cos\omega)$.

---

[14] The following expression is the result of equating the coefficients of $\cos\omega$ to zero.
[15] The following expression is the result of equating the coefficients of $\cos 2\omega$ to zero.
[16] A kind of unity, namely an alignment of three celestial bodies (for example, the Sun, Earth, and Moon) such that one body is directly between the other two, such as occurs at an eclipse (from the *Wikipedia*).

The excursions are maximum for angles $\omega$ equal to $90°, 270°$, etc., therefore, between one maximum digression to the next we have $\alpha\zeta = 180°$ and $\zeta = 86°, 53\frac{1'}{2}$: with the greatest of these digressions given by $v = b$. But in case this kind of libration[17] happens to be greater, its determination involves considerable difficulties, because, as more accurately we wish to define all the variations, less certain we would be about the remaining that we have overlooked.

———————————————

---

[17] Is the wagging of the Moon perceived by Earth-bound observers caused by changes in their perspective. It permits an observer to see slightly different halves of the surface at different times. It is similar in both cause and effect to the changes in the Moon's apparent size due to changes in distance (from the *Wikipedia*).